\documentclass[intlimits,twoside,a4paper]{article}
\usepackage[eqsecnum]{cmpj3}
\usepackage[cp1251]{inputenc}

\usepackage{graphicx}
\usepackage{amsmath}
\usepackage{natbib}

\usepackage{bm}
\usepackage{epsfig}
\newcommand{\vvec}[1]{\mathbf{#1}}

\issue{2022}{25}{2}{23302}
\doinumber{10.5488/CMP.25.23302}

\title[Swelling of asymmetric pom-pom polymers]%
{Swelling of asymmetric pom-pom polymers in dilute solutions}%

\author[K. Haydukivska, O. Kalyuzhnyi, V. Blavatska, J. Ilnytskyi]{K. Haydukivska\orcid{0000-0002-3118-7010} \refaddr{label1,label2}
\thanks{Corresponding author: \email{wja4eslawa@icmp.lviv.ua}.},
 O. Kalyuzhnyi\orcid{0000-0002-5691-3058} \refaddr{label1},
 V. Blavatska\orcid{0000-0001-6158-1636} \refaddr{label1,label3},
 J. Ilnytskyi\orcid{0000-0002-1868-5648}\refaddr{label1,label4}
}
\addresses{
\addr{label1} Institute for Condensed Matter Physics of the National Academy of Sciences of Ukraine, 1 Svientsitskii Str., 79011, Lviv, Ukraine
\addr{label2} Institute of Physics, University of Silesia, 75 Pu\l{}ku Piechoty 1, 41-500 Chorz{\'o}w, Poland
\addr{label3} Dioscuri Centre for Physics and Chemistry of Bacteria,
Institute of Physical Chemistry, Polish Academy of Sciences, 01-224 Warsaw, Poland
\addr{label4} National University Lviv Politechnic, 12 Bandera Str, 79000, Lviv, Ukraine
}

\date{Received March 28, 2022, in final form May 30, 2022}

\begin{document}

\maketitle

\begin{abstract}In this paper we continue our recent analysis [K. Haydukivska et al., J. Mol. Liq., 2021, \textbf{328}, 115456] of complex  molecules  with two branching points at both ends of the linear backbone with  $f_1$ and $f_2$ side arms starting from them,  known as the  pom-pom polymers. Here, we analyze the asymmetric case, $f_1 \neq f_2$, by applying both the analytical approach, based on the direct polymer renormalization, and computer simulations using both dissipative particle dynamics and Monte Carlo methods. We study the role played by the molecular asymmetry of average polymer conformations, considering the infinite dilution regime and good solvent conditions.The quantitative estimates are reported for the set of universal size and shape characteristics of such molecules and for their individual branches, all the functions of $f_1$ and $f_2$. In particular, we evaluate the size ratio of the gyration radii of symmetric and asymmetric pom-pom topologies with the same molecular weight and quantitatively reveal an increase of the effective size of a molecule caused by its asymmetry. We also introduce and analyse the asymmetry factor and estimate the shift of the center of mass caused by the presence of side stars, which can serve as another characteristic of the asymmetry of pom-pom structure.
	

\end{abstract}

\section{Introduction}

The study of macromolecules of a hyperbranched structure is of great interest both from both 
academic and commercial points of view. Typical examples are commercial low density
polyethylene (LDPE), which consists of linear polyethylene backbones
with attached alkyl branches~\cite{Meissner72,Mcleish95}.
The multiple long-chain branches are found to considerably increase the viscosity of the melts of such molecules due to their mutual entanglements~\cite{McLeish98}. In general, the properties of such macromolecules are determined by the competition between the  steric repulsion between the side chains and the configurational entropy of the main backbone: varying the  density of the side chains and their length leads to the change of the effective stiffness of a backbone over a wide range.

The simplest case of a polymer architecture, that correctly captures the
 characteristics of commercial long-chain
branched polymers, is the  pom-pom structure
\cite{Bishko97}.  It is assumed to consist of
a linear backbone chain with two terminal branching points
of functionalities $f_1$ and $f_2$, respectively. The
rheological properties of these structures in melts were investigated in
references~\cite{McLeish98,Graham01,Ruymbeke07,Chen11}.

In our study, we concentrated on the conformational
properties of such macromolecules in an infinite dilution regime. In our preceding work~\cite{Haydukivska21}, we studied the simplified symmetric case with $f_1=f_2$. 
The main attention was focused on a set of universal conformational
characteristics of macromolecules, which are independent of any details
of the chemical structure and are governed only by the so-called global
parameters. In particular, to estimate the impact of the complex topology of a molecule on its effective size measure in a solvent, it is useful to  consider the so-called size ratio $g$ of mean-squared gyration radii of the complex molecule $\langle R_g^2 \rangle_{{\rm complex}} $ and that of the simplest linear polymer chain
 $\langle R_g^2 \rangle_{{\rm chain}} $ of the same total molecular weight~\cite{Zimm49}. For the case of a pom-pom topology, 
in the simplest case of Gaussian polymers, when the monomer-monomer excluded volume is neglected, one has~\cite{Zimm49,Radke96}:
\begin{equation}
g_{f_1,f_2}=\frac{3(f_1^2+f_2^2)+4(f_1+f_2)+12f_1f_2+1}{(f_1+f_2+1)^2}. \label{gpom}
\end{equation}
More subtle universal characteristics, specific to branched polymers,
are the individual  branch swelling ratios $g(f)$, defined as the ratio of the averaged  gyration radii of side or backbone branches to that of a free linear chain of the same molecular weight. These ratios characterize the averaged effect of crowdedness, caused by a mutual influence of adjacent branches on their respective conformations. Applying both the theoretical approach and the computer simulations, we quantitatively described the effects of the stretch and ``compactization'' of both the symmetric pom-pom polymer and of its individual branches by means of  9 universal size ratios~\cite{Haydukivska21}. 

In the present paper, we generalize this analysis to the case of asymmetric pom-pom molecular topology and provide an impact it has on the swelling properties of such molecules in a good solvent, given by a set of the relevant size and shape ratios. We see at least two experimental situations, when such asymmetry can appear. The first one covers the case of precisely controlled synthetic protocol and intentional synthesis of an asymmetric polymer for special purpose (e.g., asymmetrical shape, specific behavior under a flow, etc.). The second one is related to the synthetic protocol, which generates some level of natural polydispersity in the functionality of the terminal beads.

The layout of the rest of the paper is as follows. Sections \ref{methods_an} and \ref{methods_num} provide a brief encounter with both analytical and numerical methods used in our study. We  introduce the continuous and the random-flight model of a polymer, describe the direct polymer renormalization approach; the numerical methods of the dissipative particle dynamics and the pivot algorithm of the lattice Monte Carlo method. In the following section \ref{results}, we report the obtained  results, that are based on the approaches described. The conclusions and an outlook are provided in section \ref{conclusions}.

\section{The methods: analytical approach}\label{methods_an}

\subsection{Continuous chain model}
Within the continuous chain model~\cite{Edwards},
pom-pom polymer is described as a set of trajectories in continuous space. Each of the trajectories has the length $L$ and is parametrized by radius vector $\vvec{r}(s)$ in $d$-dimensional space, with $s$ changing from $0$ to $L$. The Hamiltonian of this model can be written as:
\begin{eqnarray}
&&H = \frac{1}{2}\sum_{i=0}^{F-1}\,\int_0^L \rd s\,\left[\frac{\rd\vvec{r}_i(s)}{\rd s}\right]^2+\frac{u}{2}\sum_{i,j=0}^{F-1}\int_0^L \rd s'\int_0^L \rd s''\,\delta(\vvec{r}_i(s')-\vvec{r}_j(s'')),\label{H}
\end{eqnarray}
where $F=f_1+f_2+1$ is the number of trajectories, the first term in the expression~(\ref{H}) represents a connectivity of each trajectory and the second term describes a two point interaction with a coupling constant $u$ known as excluded volume interaction~\cite{desCloiseaux}.

It is a well known property of the model that all topologies are described by the same Hamiltonian~\cite{Duplantier94,Blavatska12,Blavatska15} and differ only at the level of partition function, which for the pom-pom case reads:
\begin{eqnarray}
&&Z^{\rm pom-pom}_{f_1,f_2}=\frac{1}{Z_0}\prod_{i=1}^{f_1}\prod_{j=1}^{f_2}\,\int\,D\vvec{r}(s)\,\delta(\vvec{r}_i(0)-\vvec{r}_0(0))\delta(\vvec{r}_j(0)-\vvec{r}_0(L))\,{\rm e}^{-H}.
\label{ZZ}
\end{eqnarray}
 Here, $\vvec{r}_0$ denotes a backbone 
trajectory, the two sets of delta-functions place $f_1$ and $f_2$ trajectories of side stars, correspondingly, at the terminal points of the  $0$-th trajectory, $H$ is Hamiltonian of the model given by equation~(\ref{H}), and $Z_0$ represents a partition function for a Gaussian case.

\subsection{Direct renormalization scheme}
\label{Met}
Observables calculated within the frames of continuous chain model depend on the chain length $L$ and diverge in the limit of $L\rightarrow \infty$. These divergences should be removed in order to receive  universal values for the observables in question. In this work we use the des Cloiseaux's direct renormalization scheme~\cite{desCloiseaux}, where a set of renormalization factors are introduced, that are directly connected to the physical quantities for which the divergencies should be removed.

Different topologies are introduced into the model through partition function, and the Hamiltonian of the model contains only interactions. Therefore, the fixed points of the renormalization scheme, once calculated for a particular Hamiltonian, do not depend on the topology of the molecule under consideration. Consequently, it suffices to evaluate the fixed points of the model for the simplest case.

Starting with the calculation of the partition function of two interacting polymers $Z(L,L)$, the method then introduces the renormalization factors that are connected with the number of allowed trajectories (the partition function of a single chain) $[Z(L,u_0)]^{-2}$, and with its characteristic size which  is represented by the end-to-end distance $\langle R_e^2\rangle$ or, more precisely, a swelling factor $\chi_0(L,\{x_0\})$, defined as $\chi_0=\langle R_e^2\rangle/L$. Those functions allow one to introduce a renormalized coupling constant:
\begin{eqnarray}
u_R(u_0)= - [Z(L,u_0)]^{-2}Z(L,L)[2\piup \chi_0(L,u_0)]^{-2+\epsilon/2},
\end{eqnarray}
with $\epsilon=4-d$ being a deviation from the upper critical dimension for the coupling constant $u_0$.

In the limit of infinitely long chains, the dimensionless coupling constant $u_0=u(2\piup)^{-d/2}L^{2-d/2}$ diverges unlike the renormalized one $u_R$, which in the same limit reaches a fixed value:
\begin{eqnarray}
\lim_{L\to\infty} u_{R}(u_0)=u_R^*.
\end{eqnarray}
In practice, the value $u_R^*$ is calculated from the equation:
\begin{eqnarray}
\beta_{u_R}=2L\frac{\partial u_R(u_0)}{\partial L}=0.
\end{eqnarray}
For the model under consideration, those solutions are well known, and in the first order of the $\epsilon$-expansion read:\cite{desCloiseaux}:
\begin{eqnarray}
&& {\rm {Gaussian:}} \qquad u^*_{R}=0, \label{FPG}\\
&& { \rm {Pure:}} \qquad u^*_{R}={\epsilon}/{8}, \label{FPP}
 \end{eqnarray}
where the first one describes a Gaussian case, and the second one is a coupling constant for the model with excluded volume interaction.

\subsection{Random-flight model} \label{flight}

Within the formalism of the random flight model~\cite{Yamakawa}, every linear sub-chain of $N$ monomers within a complex polymer structure is considered as a sequence of 
$N$ bond vectors $\sigma_k$, with $0\leqslant k \leqslant N$ with the origin denoted as $0$. We assume it to coincide with the center of the central (backbone) chain.
Let us denote by ${\vvec{R}_{\rm cl}}_n$ and ${\vvec{R}_{\rm cr}}_n$ the position vectors
of segments in the central chain  to the right and to the left from the reference point (considered as $0$), correspondingly. They can be expressed as sums 
over the set of corresponding bond vectors:
\begin{eqnarray}
&&{\vvec{R}_{\rm cl}}_n=\sum_{k=1}^n {\boldsymbol{\sigma}_{\rm  cl}}_k, \qquad {\vvec{R}_{\rm cr}}_n=\sum_{k=1}^n {\boldsymbol{\sigma}_{\rm  cr}}_k,\qquad 0\leqslant n \leqslant \frac{N}{2} \label{centralR}.
\end{eqnarray} 
Similarly, let
 $\vvec{R}^{i}_{{\rm pl}_n}$ and $\vvec{R}^{i}_{{\rm pr}_n}$ be the position
 vectors of monomers in the left and right side stars (poms), correspondingly, so that
\begin{eqnarray}
&&{\vvec{R}_{\rm pl}}_n^{i}=\sum_{k=1}^{N/2}{\boldsymbol{\sigma}_{\rm  cr}}_k+\sum_{k=1}^n {\boldsymbol{\sigma}_{\rm  cr}}_k^{i}, \qquad i=1,\ldots,f_1, \label{pomR1}\\
&& {\vvec{R}_{\rm pr}}_n^{i}=\sum_{k=1}^{N/2} {\boldsymbol{\sigma}_{\rm  cr}}_k+\sum_{k=1}^n {\boldsymbol{\sigma}_{\rm  cr}}_k^{i}, \qquad i=1,\ldots,f_2, \qquad \label{pomR2}\frac{N}{2}\leqslant n \leqslant N+\frac{N}{2}.\nonumber
\end{eqnarray} 
All the bond vectors are assumed to have an equal length $l$ and are connected to each other by unrestricted joints. The direction of
each bond is random and independent of the directions of its neighbours.  Thus, performing the averaging over an ensemble of possible bond configurations, we have
\begin{eqnarray}
&&\langle \boldsymbol{\sigma}_k \rangle=0, \qquad \langle \boldsymbol{\sigma}_k^2 \rangle=l^2, \label{sigma1}\\
&&\langle {\boldsymbol{\sigma}^i_{\rm  a}}_k {\boldsymbol{\sigma}^j_{\rm  b}}_m  \rangle=l^2\delta_{ij}\delta_{km}\delta_{ab}, \qquad \{ {\rm a,b}\} \in \{ {\rm cl,cr,pl,pr} \} \label{sigma2},
\end{eqnarray}  
with $\delta$ being the Kronecker delta symbol. In what follows, we take $l=1$.

Note that the polymer molecule  in this model is assumed to capture Gaussian statistics, i.e., the excluded volume (self-avoidance of segments)
 is neglected.

\section{The methods: numerical approaches}\label{methods_num}
\subsection{Lattice model and the pivot algorithm}

\begin{figure}[t!]
\begin{center}
\includegraphics[width=43mm]{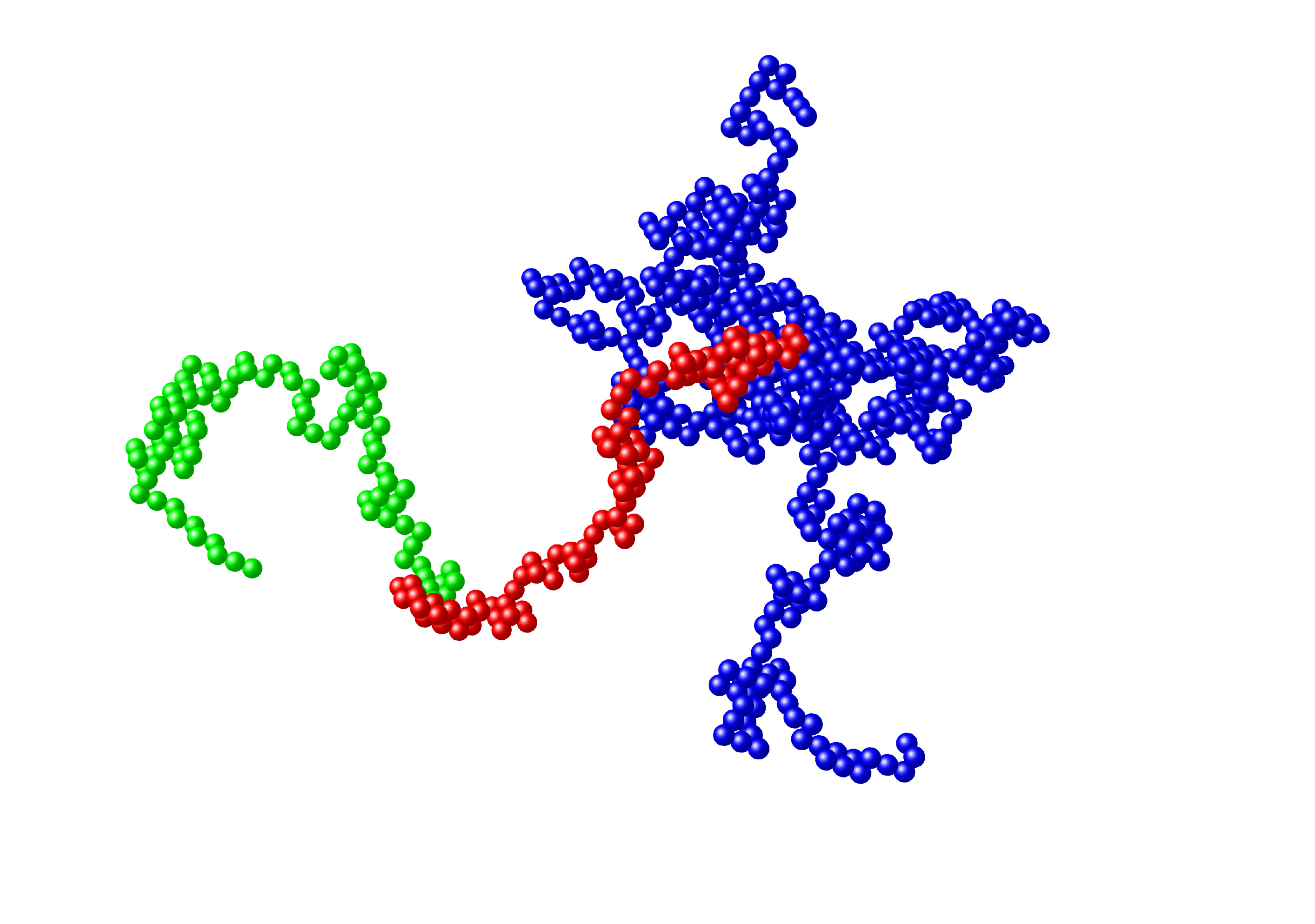}\includegraphics[width=43mm]{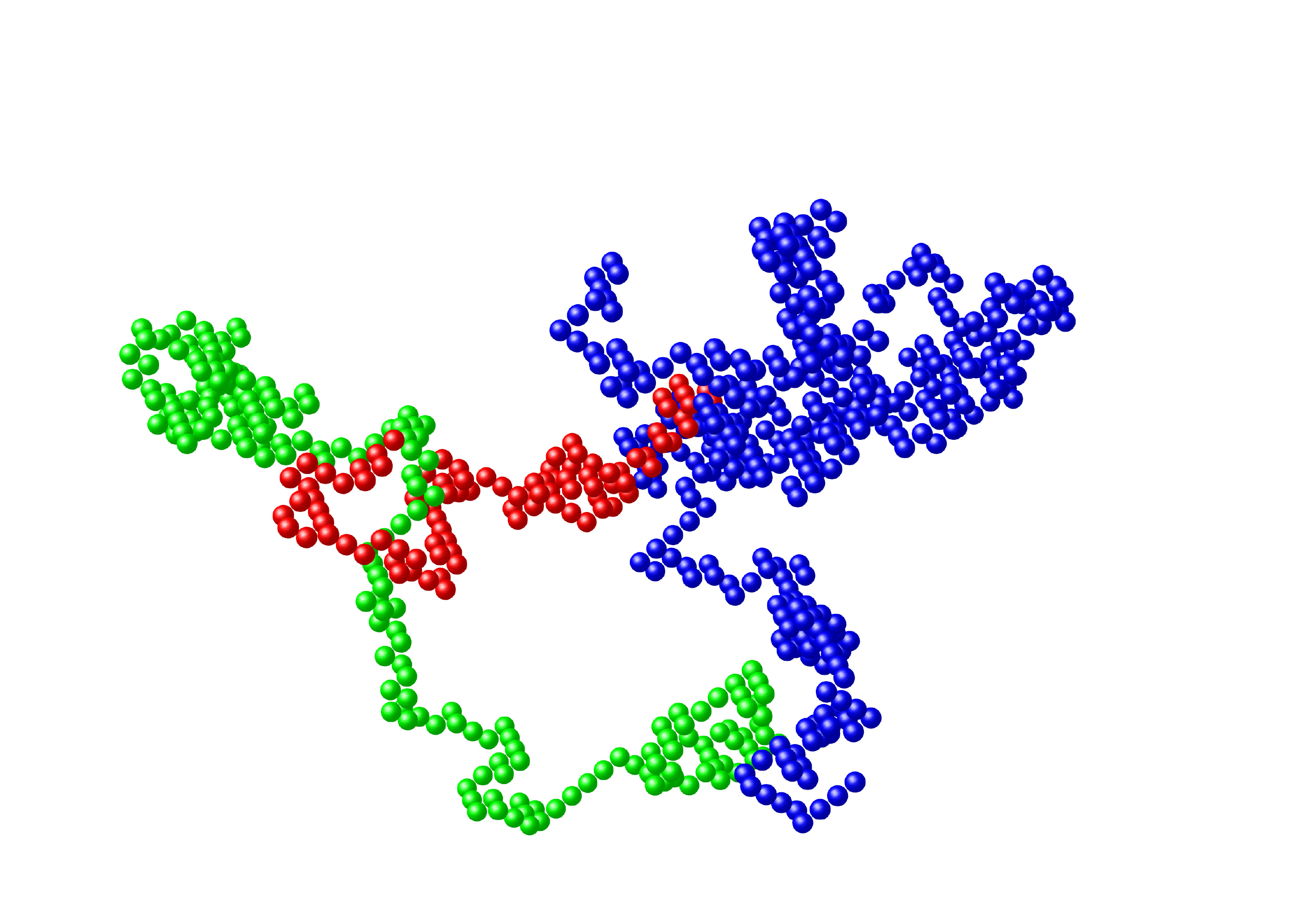}
\caption{ \label{fig:5} (Colour online) Snapshots of the pom-pom conformations for different values of $f_1$ and $f_2$ after $2000 f_1 f_2$ Monte Carlo steps using the pivot algorithm. On the left: $f_1=1$ and $f_2=5$; on the right:~$f_1=2$ and $f_2=5$.}
\end{center}
\end{figure}

Numerical simulations for a pom-pom polymer are performed within the lattice model of self-avoiding walk (SAW) on a simple cubic lattice via pivot algorithm~\cite{Clisby10,Madras88}. Each trajectory is forbidden to visit the site occupied by itself or by the other trajectory. In total, we consider $f_1+f_2+1$ walks, so that $f_1+1$ of them start at the origin and $f_2$ --- at the end of the walk numbered as  $0$-th. We consider both branching parameters $f_1$ and $f_2$ to vary from $1$ to $13$ with all walks being of the same length top bounded by $N=100$ steps, making a total amount of steps in the pom-pom to be $N(f_1+f_2+1)$. Then, the  pivot algorithm is applied again to receive a new configuration. We start from the initial configuration as a set of straight lines. Then, a randomly chosen transformation is applied to the part of a walk. If the new trajectory is acceptable under the self-avoidance rules, the new conformation is accepted; if not --- it is rejected and the previous one is accounted for one more time. Then, another pivot operation is performed at a new randomly chosen site. The first $20N(f_1+f_2+1)$ operations are rejected for the equilibration, and the next $10^5$ are used for calculation of the observables. All observables are evaluated as the mean arithmetic average over the set of structures obtained by means of simulations. Numerical modelling is performed for $N$ varying from $10$ to $100$ with the step of $10$. Final values are received by application of the finite size approximation $g_x=g_{x,\infty}+B/N$, with $B$ being a constant. The snapshots of the obtained conformations are illustrated in figure~\ref{fig:5}.

\subsection{Dissipative particle dynamics simulations}

Another numeric approach used in this study is the dissipative particle dynamics (DPD)~\cite{Groot1997}.  In this
method, the polymer and solvent molecules are modelled as soft beads
of equal size, each representing a group of atoms. For this reason, the interaction potential between two beads are, in general, density and temperature dependent. Different parametrization strategies are possible and in this paper we closely followed the original parametrization by Groot and Warren based on matching the model compressibility to that of water~\cite{Groot1997}. The length-scale is
given by the diameter of a soft bead, and the energy unit is
$k_{\rm B}T=1$, where $k_{\rm B}$ is the Boltzmann constant and $T$ is the
temperature. Monomers in a polymer chain are bonded via harmonic
springs resulting in a force:
\begin{equation}\label{FB}
\vvec{F}^B_{ij} = -k\vvec{x}_{ij}\,,
\end{equation}
where $k$ is the spring constant, and
$\vvec{x}_{ij}=\vvec{x}_i-\vvec{x}_j$, $\vvec{x}_i$ and $\vvec{x}_j$
are the coordinates of $i$-th and $j$-th bead, respectively. The
non-bonding force $\vvec{F}_{ij}$ acting on the $i$-th bead as a result of its interaction with
its $j$-th counterpart is expressed as a sum of three contributions
\begin{equation}
\vvec{F}_{ij} = \vvec{F}^{\mathrm{C}}_{ij} + \vvec{F}^{\mathrm{D}}_{ij}
+ \vvec{F}^{\mathrm{R}}_{ij}\,,
\end{equation}
where $\vvec{F}^{\mathrm{C}}_{ij}$ is the conservative force responsible for the repulsion between the beads,
$\vvec{F}^{\mathrm{D}}_{ij}$ is the dissipative force mimicking the friction between them and the random force
$\vvec{F}^{\mathrm{R}}_{ij}$  works in pair with a dissipative force
to thermostat the system. The expression for all these three
contributions are given below~\cite{Groot1997}
\begin{equation}\label{FC}
\vvec{F}^{\mathrm{C}}_{ij} =
\left\{
\begin{array}{ll}
a(1-x_{ij})\displaystyle\frac{\vvec{x}_{ij}}{x_{ij}}, & x_{ij}<1,\\
0,                       & x_{ij}\geqslant 1,
\end{array}
\right.
\end{equation}
\begin{equation}\label{FD}
\vvec{F}^{\mathrm{D}}_{ij} = -\gamma
w^{\mathrm{D}}(x_{ij})(\vvec{x}_{ij}\cdot\vvec{v}_{ij})\frac{\vvec{x}_{ij}}{x^2_{ij}},
\end{equation}
\begin{equation}\label{FR}
\vvec{F}^{\mathrm{R}}_{ij} = \sigma
w^{\mathrm{R}}(x_{ij})\theta_{ij}\Delta t^{-1/2}\frac{\vvec{x}_{ij}}{x_{ij}},
\end{equation}
where $a$ is the amplitude for the conservative repulsive force,
$x_{ij}=|\vvec{x}_{ij}|$,  $\vvec{v}_{ij}=\vvec{v}_{i}-\vvec{v}_{j}$, 
$\vvec{v}_{i}$ is the velocity of the $i$-th bead. The dissipative
force has an amplitude $\gamma$ and decays with distance according
to the weight function $w^{\mathrm{D}}(x_{ij})$. The amplitude for
the random force is $\sigma$ and the respective weight function is
$w^{\mathrm{R}}(x_{ij})$. $\theta_{ij}$ is the Gaussian random
variable. As was shown by Espa\~{n}ol and Warren~\cite{Espanol1995},
to satisfy the detailed balance requirement, the amplitudes and
weight functions for the dissipative and random forces should be
interrelated: $\sigma^2=2\gamma$ (we choose $\gamma=6.75$, $\sigma=\sqrt{2\gamma}=3.67$ here)
and $w^{\mathrm{D}}(x_{ij})=\left[w^{\mathrm{R}}(x_{ij})\right]^2$. Here,
we use the weight functions quadratically decaying with the distance:
\begin{equation}
w^{\mathrm{D}}(x_{ij})=\left[w^{\mathrm{R}}(x_{ij})\right]^2
=\left\{
\begin{array}{ll}
(1-x_{ij})^2, & x_{ij} < 1,\\
0, & x_{ij} \geqslant 1.
\end{array}
\right.
\end{equation}
Following parametrization of Groot and Warren~\cite{Groot1997}, the reduced number density of the system is set at $\rho^{*} = N/V=3$, where $N$ is the total number of beads
(pom-pom polymer and a solvent) in a system and $V$ is the volume of a simulated system. We consider the case of a single pom-pom polymer within a simulation box, which reproduces the conditions of an infinite dilution. To avoid self-interaction between periodic images of a pom-pom molecule, the linear dimension of a cubic simulation box $L$ was chosen accordingly. In particular, we used the expression: $L\approx 1.75 R^*_g$, where $R^*_g=(3N_f-1)^{\nu}$ estimates the radius of gyration of the longest linear sub-chain of a pom-pom of length $3N_f$ and bond length of $1$ in a good solvent regime. Here $N_f$ is the number of beads in a single arm and $\nu=0.59$ is the Flory exponent. 

The total number of branches was fixed at $F=f_1+f_2 + 1 = 15$ in all cases.  This number of branches was chosen to show the effect of asymmetry and to avoid the effect of crowdedness which has been studied in our previous work~\cite{Kalyuzhnyi2019}. Two cases of the arm length, $N_f = 8$ and $N_f = 16$ were examined, similarly to the case of a symmetric pom-pom polymer~\cite{Haydukivska21}, the results were found to agree well.  In work~\cite{Groot1997} it was shown that the reduced density is coupled to the amplitude $a$ of a repulsive force in equation~(\ref{FC}). Following this paper, we set time-step $\Delta t = 0.04$ and $a=25$ for all combinations of the interacting beads: polymer-polymer, solvent-solvent and polymer-solvent.  

The first $2\cdot10^6$ steps are allowed for the system equilibration and are skipped from the analysis. The productive runs span from the step $2\cdot10^6$ up to the maximal simulation step in each case (typically $8\cdot10^6$). The error estimates are made by splitting the whole productive run in four equal pieces and by evaluating the partial averages~$A_i$, $i=1-4$ for a given property $A$ in each of them. The final average is $\langle A \rangle=\frac{1}{4}{\sum_{i=1}^{4}A_i}$, whereas the conservative estimate for the standard error is given by
\begin{displaymath}
e(A)=\left[  \frac{1}{4}\displaystyle{\sum_{i=1}^{4}(A_i-\langle A \rangle)^2}
\right]^{{1}/{2}}.
\end{displaymath}

\section{Results}\label{results}

\subsection{Partition function calculation}

\begin{figure}[b!]
\begin{center}
\includegraphics[width=73mm]{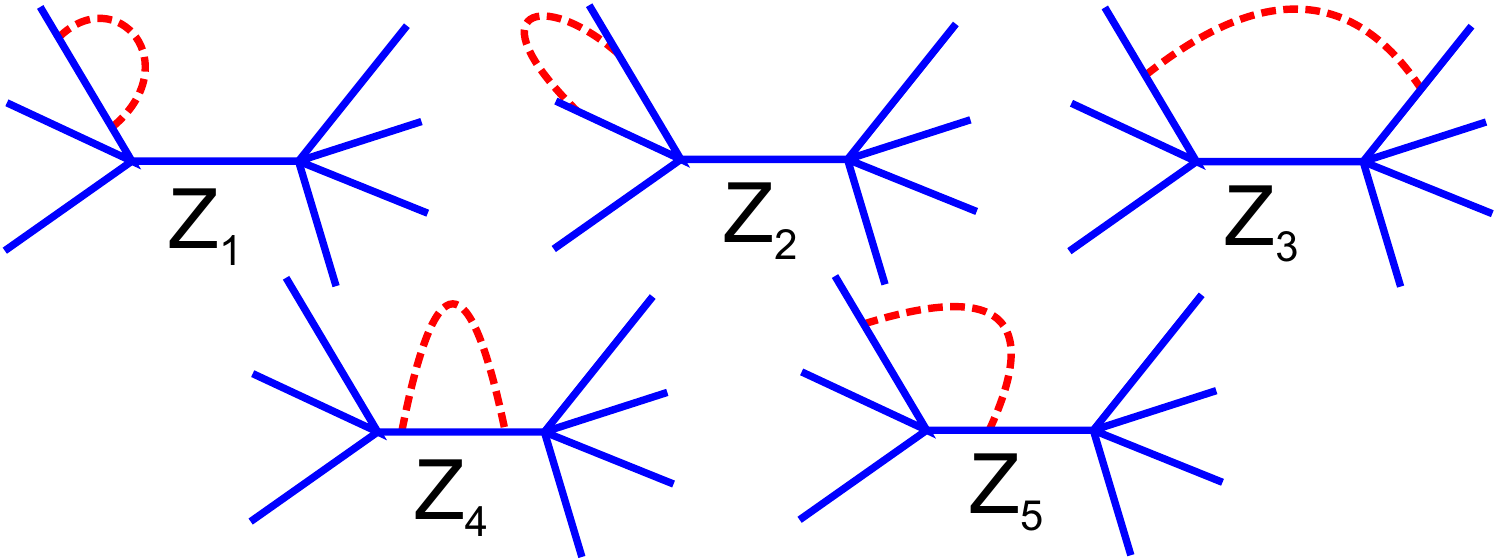}
\caption{ \label{fig:2} (Colour online) Diagrammatic representation of the partition function terms in the one-loop approximation. The solid lines represent schematically a polymer path of the length $L$ each, and a dash line represents the two monomer excluded volume interaction. The diagrams are enumerated by a label $Z_i$, $i=1,\ldots,5$.
}
\end{center}
\end{figure}

We start with the evaluation of the expression for the partition function~(\ref{ZZ}) as a series in a coupling constant~$u$, where the zero order term corresponds to the unperturbed Gaussian case. We restrict the series to the linear term in $u$, the so-called one loop approximation:
\begin{align}
Z^{\rm pom-pom}_{f_1,f_2}&=\frac{1}{Z_0}\prod_{i=1}^{f_1}\prod_{j=1}^{f_2}\,\int\,D\vvec{r}(s)\,\delta(\vvec{r}_i(0)-\vvec{r}_0(0))\nonumber\\
&\times\,\delta(\vvec{r}_j(0)-\vvec{r}_0(L))\left\{\exp{\left[-\frac{1}{2}\sum_{i=0}^{F-1}\,\int_0^L \rd s\,\left(\frac{\rd\vvec{r}_i(s)}{\rd s}\right)^2\right]}-\right.\nonumber\\
&-\frac{u}{2}\sum_{i,j=0}^{F-1}\int_0^L\rd s'\int_0^L \rd s''\,\delta(\vvec{r}_i(s')-\vvec{r}_j(s''))
\left.\,\exp{\left[-\frac{1}{2}\sum_{i=0}^{F-1}\,\int_0^L \rd s\,\left(\frac{\rd \vvec{r}_i(s)}{\rd s}\right)^2\right]}\right\},
\label{ZZ1}
\end{align}
where
\begin{eqnarray}
Z_0&=&\prod_{i=1}^{f_1}\prod_{j=1}^{f_2}\,\int\,D\vvec{r}(s)\,\delta(\vvec{r}_i(0)-\vvec{r}_0(0))\delta(\vvec{r}_j(0)-\vvec{r}_0(L))\exp{\left\{-\frac{1}{2}\sum_{i=0}^{F-1}\,\int_0^L \rd s\,\left[\frac{\rd \vvec{r}_i(s)}{\rd s}\right]^2\right\}}\nonumber
\end{eqnarray}
is the Gaussian result for the partition function of a pom-pom polymer, in which case the excluded volume interactions are neglected. In order to calculate the contribution from the second term in (\ref{ZZ1}), we consider a set of diagrams presented in figure~\ref{fig:2}. It is important to note that the diagram $Z_1$ describes the excluded volume interaction within a single trajectory and needs to be taken into account with the pre-factor $f_1+f_2+1$,
diagram $Z_2$ corresponds to the contributions from all the interactions between two chains that are separated by one branching point, and contains a pre-factor $f_1+f_2+[f_1(f_1-1)+f_2(f_2-1)]/2$, and the last diagram corresponds to the interactions between the chains separated by both branching points and has a pre-factor $f_1f_2$.

Using a Fourier transform presentation of $\delta$-function:
\begin{eqnarray}
\delta (\vvec{r}_i(s')-\vvec{r}_j(s'')) =\frac{1}{(2\piup)^{d}} \int {\rm d}\vvec{p}_u\, \rm{exp}\{-\ri\vvec{p}_u[\vvec{r}_i(s')-\vvec{r}_j(s'')]\},
\end{eqnarray}
and calculating the contributions of all diagrams, we obtain the following expressions:
\begin{eqnarray}
&&Z_1=Z_4=\frac{u(2\piup)^{-d/2}L^{2-d/2}}{(1-d/2)(2-d/2)},
\end{eqnarray}
\begin{eqnarray}
&&Z_2=Z_5=\frac{u(2\piup)^{-d/2}L^{2-d/2}(2^{2-d/2}-2)}{(1-d/2)(2-d/2)},
\end{eqnarray}
\begin{eqnarray}
&&Z_3=\frac{u(2\piup)^{-d/2}L^{2-d/2}(3^{2-d/2}-2\,2^{2-d/2}+1)}{(1-d/2)(2-d/2)}.
\end{eqnarray}
As already mentioned in the previous section, we consider all the observables in the first order of the $\epsilon$-expansion. Therefore, the expressions for the diagrams read:
\begin{eqnarray}
&&Z_1=-{2}/{\epsilon}-1,\\
&&Z_2=-{2}/{\epsilon}-1-\ln2,\\
&&Z_3=2\ln2-\ln3.
\end{eqnarray}
Getting all terms together, we obtain the expression for the partition function of a pom-pom molecule in one-loop approximation
\begin{align}
Z&=1-u_0\biggl\{\frac{f_1(f_1-1)+f_2(f_2-1)-2}{\epsilon}+\left[f_1+f_2+\frac{f_1(f_1-1)+f_2(f_2-1)}{2}\right] [1-\ln2]\biggr.\nonumber\\
&\biggl.+f_1f_2 [2\ln2-\ln3]-f_1-f_2-1\biggr\},
\label{ZZf}
\end{align}
where $u_0=u(2\piup)^{-d/2}L^{2-d/2}$ is a dimensionless coupling constant.

All the observables calculated below are averaged over an ensemble of all possible conformations with the partition function (\ref{ZZf}):
\begin{eqnarray}
&&\langle (\ldots) \rangle = \frac{1}{{ Z^{\rm pom-pom}_{f_1,f_2}}}\prod_{i=1}^{f_1}\prod_{j=1}^{f_2}\,\int\,D\vvec{r}(s)\delta(\vvec{r}_i(0)-\vvec{r}_0(0))\delta(\vvec{r}_j(0)-\vvec{r}_0(L))\,{\rm e}^{-H}(\ldots).
\end{eqnarray}

\subsection{Size characteristics}

\begin{figure}[t!]
\begin{center}
\includegraphics[width=73mm]{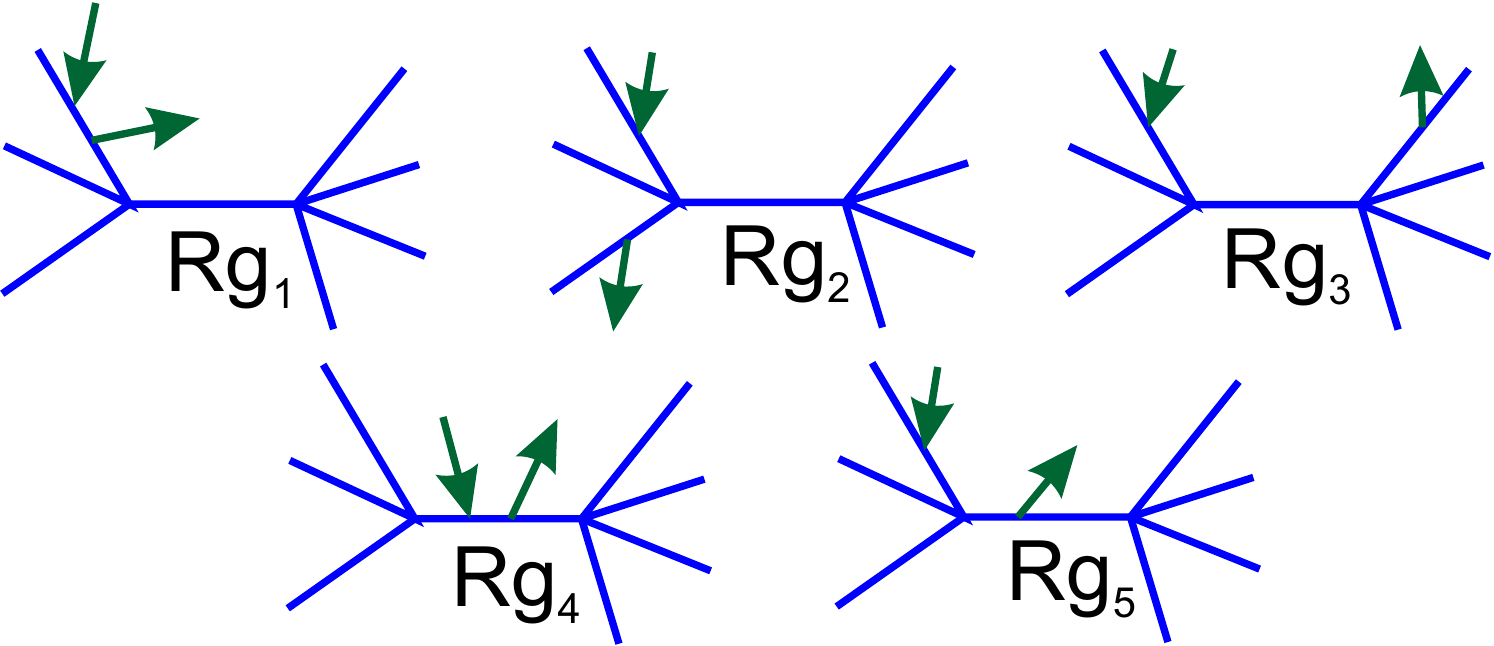}
\caption{ \label{fig:3} (Colour online) Diagrammatic representation of the $\xi(\vvec{k})$ terms in the  Gaussian approximation. The solid lines schematically represent polymer paths of the lengths $L$, and arrows represent the so-called restriction points $s_1$ and $s_2$.  
The diagrams are enumerated by a label ${\rm Rg}_i$, $i=1,\ldots,5$.
}
\end{center}
\end{figure}
A set of specific size characteristics can be defined for the pom-pom polymer. Some of them were introduced for a symmetric case in our previous study~\cite{Haydukivska21}. Here, we add a number of new characteristics that allow us to describe the asymmetry of pom-pom molecule when $f_1\neq f_2$. The gyration radius of the structure is defined as:
\begin{eqnarray}
&&{\langle {R^2_{g}}\rangle} = \frac{1}{2L^2(F)^2} \sum_{i,j=0}^{F-1}\int_0^L\int_0^L \rd s_1\,\rd s_2 \langle(\vvec{r}_i(s_2)-\vvec{r}_j(s_1))^2\rangle,
\end{eqnarray}
where $s_2$ and $s_1$ are the ``monomer numbers'' along the chain, such that the average square distance is calculated between them. They are often called restriction points. In order to calculate this quantity we use the identity:
\begin{eqnarray}
&&\overline{\langle(\vvec{r}_i(s_2)-\vvec{r}_j(s_1))^2\rangle} = - 2 \frac{\rd}{\rd|\vvec{k}|^2}\xi(\vvec{k})_{\vvec{k}=0},\nonumber\\
&&\xi(\vvec{k})\equiv\langle{\rm{exp}}\{-\ri\vvec{k}[\vvec{r}_i(s_2)-\vvec{r}_j(s_1)]\}\rangle ,\label{identity}
\end{eqnarray}
and evaluate $\xi(\vvec{k})$ in the path integration approach.
In calculations of the contributions into $\xi(\vvec{k})$, it is convenient to use the diagrammatic presentation, as given in
figure~\ref{fig:3} for the Gaussian chain.

It is interesting to note that diagrams in figures~\ref{fig:2} and \ref{fig:3} look similar with the only difference: in the former we consider them with interactions and in the latter one with the restriction points, though the pre-factor will be the same and the expression for the gyration radius in the Gaussian approximation reads:
\begin{eqnarray}
&&\langle R^2_{g,{\rm pom-pom}}\rangle_0 =\frac{dL}{6F^2}\left(3f_1^2+3f_2^2+4f_1++4f_2+12f_1f_2+1\right).\nonumber
\end{eqnarray}
Considering all possible combinations for diagrams in figures~\ref{fig:2} and \ref{fig:3}, a contribution to $\xi(\vvec{k})$ the one-loop approximation is obtained:
\begin{eqnarray}
&&\langle R^2_{g,{\rm pom-pom}}\rangle=\langle R^2_{g,{\rm pom-pom}}\rangle_0 \left[1+u_0C_g(f_1,f_2,d)\right].
\label{RG}
\end{eqnarray}
An expression for $C_g$ is provided in the Appendix in expression \ref{A_rg}. In the case of $f_1=f_2$ we recover our previous result for the symmetric case~\cite{Haydukivska21}.

We introduce two other characteristics that quantify the asymmetry of the molecular conformation: the average squared distances from a monomer to the first and to the second branching point, respectively:
\begin{eqnarray}
{\langle {r_1^2}\rangle} =\frac{1}{FL}\sum_{i=0}^{F-1}\int_0^L \,\rd s{\langle(\vvec{r}_i(s)-\vvec{r}_0(0))^2\rangle},\label{r1}\\
{\langle {r^2_2}\rangle} =\frac{1}{FL}\sum_{i=0}^{F-1}\int_0^L \,\rd s{\langle(\vvec{r}_i(s)-\vvec{r}_0(L))^2\rangle}.\label{r2}
\end{eqnarray}
In the process of evaluating these parameters, we again use the diagrammatic technique and consider a set of diagrams similar to those for the gyration radius but with an additional  restriction: points are permanently fixed at the branching point. The expressions in the  Gaussian approximation read:
\begin{eqnarray}
&&{\langle {r_1^2}\rangle} =\frac{dL(f_1+1+3f_2)}{2 F},\label{r1g}\\
&&{\langle {r^2_2}\rangle} =\frac{dL(f_2+1+3f_1)}{2 F}.\label{r2g}
\end{eqnarray}
Similarly, the expressions in the one loop approximation can be presented in a general form:
\begin{eqnarray}
&&{\langle {r_1^2}\rangle} =\frac{dL(f_1+1+3f_2)}{2 F} \left[1+u_0C_1(f_1,f_2,d)\right],\label{r_1}\\
&&{\langle {r^2_2}\rangle} =\frac{dL(f_2+1+3f_1)}{2 F} \left[1+u_0C_2(f_1,f_2,d)\right],
\label{r_2}\end{eqnarray}
with $C_1(f_1,f_2,d)$ and $C_2(f_1,f_2,d)$ being the functions of the space dimension and topology. $\epsilon$-expansions for $\langle {r_1^2}\rangle$ and $\langle {r_2^2}\rangle$ are provided in the Appendix in expressions (\ref{A_C1}) and (\ref{A_C2}), respectively.

\subsection{Universal size ratios}

To describe the impact of the effects of asymmetry on the size and shape of pom-pom structure, we consider a set of size ratios. We start by considering the ratios that compare the size of the asymmetric pom-pom to the other molecular topologies: a linear polymer chain, a star,  and a symmetric pom-pom of the same total molecular weight, defined as:
\begin{eqnarray}
&&g_c=\frac{\langle R^2_{g,{\rm pom-pom}}\rangle}{\langle R^2_{g,{\rm chain}}\rangle},\label{gc}\\
&&g_s=\frac{\langle R^2_{g,{\rm star}}\rangle}{\langle R^2_{g,{\rm pom-pom}}\rangle},\label{gs}\\
&&g_p=\frac{\langle R^2_{g,{\rm pom-pom}}\rangle}{\langle R^2_{g,{\rm pom-pom,sym}}\rangle},\label{gp}
\end{eqnarray}
respectively.
Within the renormalization group approach, these quantities are presented as the series in $\epsilon=4-d$
\begin{eqnarray}
&&\langle R^2_{g,{\rm chain}}\rangle=\frac{LdF}{6}\left[1+\frac{2 u_0}{\epsilon}+u_0 A_g(0,0)\right],\label{chain}\\
&&\langle R^2_{g,{\rm star}}\rangle=\frac{Ld(3F-2)}{6F}\left[1+\frac{2 u_0}{\epsilon}+u_0 A_g(F-1,0)\right],\label{star}\\
&&\langle R^2_{g,{\rm pom-pom,sym}}\rangle=\frac{Ld(18f^2+8f+1)}{6F^2}\left[1+\frac{2 u_0}{\epsilon}+u_0 A_g(f,f)\right],\label{pps}\\
&&\langle R^2_{g,{\rm pom-pom}}\rangle=\langle R^2_{g,{\rm pom-pom}}\rangle_0\left[1+\frac{2 u_0}{\epsilon}+u_0 A_g(f_1,f_2)\right],\label{ppa}
\end{eqnarray}
with the expressions for $A_g(0,0), A_g(f,f), A_g(f_1,f_2)$ provided in the Appendix [expression~(\ref{A_A_g})].
Note that the first two expressions above reproduce the well known results for the gyration radius of a linear chain~\cite{desCloiseaux} and of a star~\cite{Blavatska12}.

Thus, the final expressions for the size ratios (\ref{gc}, \ref{gs}) and (\ref{gp}) read:
\begin{eqnarray}
&&g_c=\frac{6\langle R^2_{g,{\rm pom-pom}}\rangle}{LdF}\left\{1+\frac{\epsilon}{8}[A_g(f_1,f_2)-A_g(0,0)]\right\},\label{gc1}\\
&&g_s=\frac{Ld(3F-2)}{6F\langle R^2_{g,{\rm pom-pom}}\rangle}\left\{1+\frac{\epsilon}{8}[A_g(F-1,0)-A_g(f_1,f_2)]\right\},\label{gs1}\\
&&g_p=\frac{6F^2\langle R^2_{g,{\rm pom-pom}}\rangle}{Ld(18f^2+8f+1)}\left\{1+\frac{\epsilon}{8}[A_g(f_1,f_2)-A_g(f,f)]\right\},\label{gp1}
\end{eqnarray}
where $u_0$ is replaced by the fixed point (\ref{FPP}).

Let us note that the expressions (\ref{gc1}, \ref{gs1}) and (\ref{gp1}) contain the first order terms of the perturbation theory only and, therefore, provide rather qualitative than the quantitative information. It is, however, known that the renormalization group approach requires at least the second order terms in $\epsilon$ to get the quantitatively sound values. This requires the evaluation of the terms up to $u_0^2$, since $u_0\sim \epsilon$. These calculations for the case of branched structures are very challenging, in particular, since  the total number of diagrams grows exponentially and leads to cumbersome expressions that are difficult to handle analytically. Because of this, it is practical to use the approximation proposed in reference~\cite{Douglas84},
based on a general idea of presenting the gyration radius in a form:
\begin{equation}
\langle R_g^2 \rangle = \langle R_g^2 \rangle_0 \left({2\piup N}/{\Lambda}\right)^{2\nu(\eta)-1}f_p(\eta),\label{SF}
\end{equation}
with $N$ being the number of monomers, $\Lambda$ is a coarse-grained length scale. Parameter $\eta$ here defines the transition from the Gaussian case at $\eta=0$ [$\nu(\eta)=1/2$ and $f_p(\eta)=1$] to the case of a polymer with an excluded volume interaction at $\eta\rightarrow \infty $ [$\nu(\eta)=\nu$ and $f_p(\eta)=1+a$]. Since only the factors $\langle R_g^2 \rangle_0$ and $f_p(\eta)$ depend on the polymer topology, the size ratio can be presented as:
\begin{equation}
g_x = \frac{\langle R_{g,1}^2 \rangle_0}{\langle R_{g,2}^2 \rangle_0}\frac{1+a_1}{1+a_2},
\end{equation}
with $ {\langle R_{g,1}^2 \rangle_0}/{\langle R_{g,2}^2 \rangle_0}$ being the ratio between topologies ``$1$'' and ``$2$'' in Gaussian approximation. And the parameters $a_1$, $a_2$ are calculated for these topologies using an expression:
\begin{equation}
a=\frac{3}{32}\frac{C_g(f_1,f_2,d=3)}{F^{{(4-d)}/{2}}}-\frac{1}{4}. \label {a}
\end{equation}
This approximation allows us to partially account for the higher orders in $\epsilon$ in the $C_g$ expansion, resulting in such terms to be added to  equations~(\ref{gc1}, \ref{gs1})  and (\ref{gp1}).

The ratios (\ref{gc}, \ref{gs}) and (\ref{gp}) allow us to examine how the asymmetricity of a pom-pom structure affects its overall shape as compared to the cases of a linear chain, star-polymer and a symmetric pom-pom structure. It would also be informative to introduce some specific characteristic, which addresses the shape asymmetry explicitly.  For that purpose, we suggest the following combination, which can be termed as the asymmetry factor:
\begin{eqnarray}
g_{\rm{asym}}=\left|\frac{\langle r_1^2\rangle}{ \langle r_2^2\rangle}-\frac{\langle r_2^2\rangle}{ \langle r_1^2\rangle} \right|,\label{gasym}
\end{eqnarray}
with $\langle r_1^2\rangle$ and $\langle r_2^2\rangle$
given by equations~(\ref{r1}) and (\ref{r2}). It is strictly zero for the symmetric case, $f_1=f_2$, and its maximum value is reached for the most asymmetric cases, $f_1=1$ and $f_2=F-2$, or vice versa. In the Gaussian approximation, the corresponding expression reads:
\begin{eqnarray}
g_{\rm{asym}}=\left|\frac{[4(2f_1+1+2f_2)](f_2-f_1)}{(f_2+1+3f_1)(f_1+1+3f_2)} \right|.
\end{eqnarray}

\subsection{Center of mass}

\begin{figure}[t]
	\begin{center}
		\includegraphics[width=73mm,angle=270]{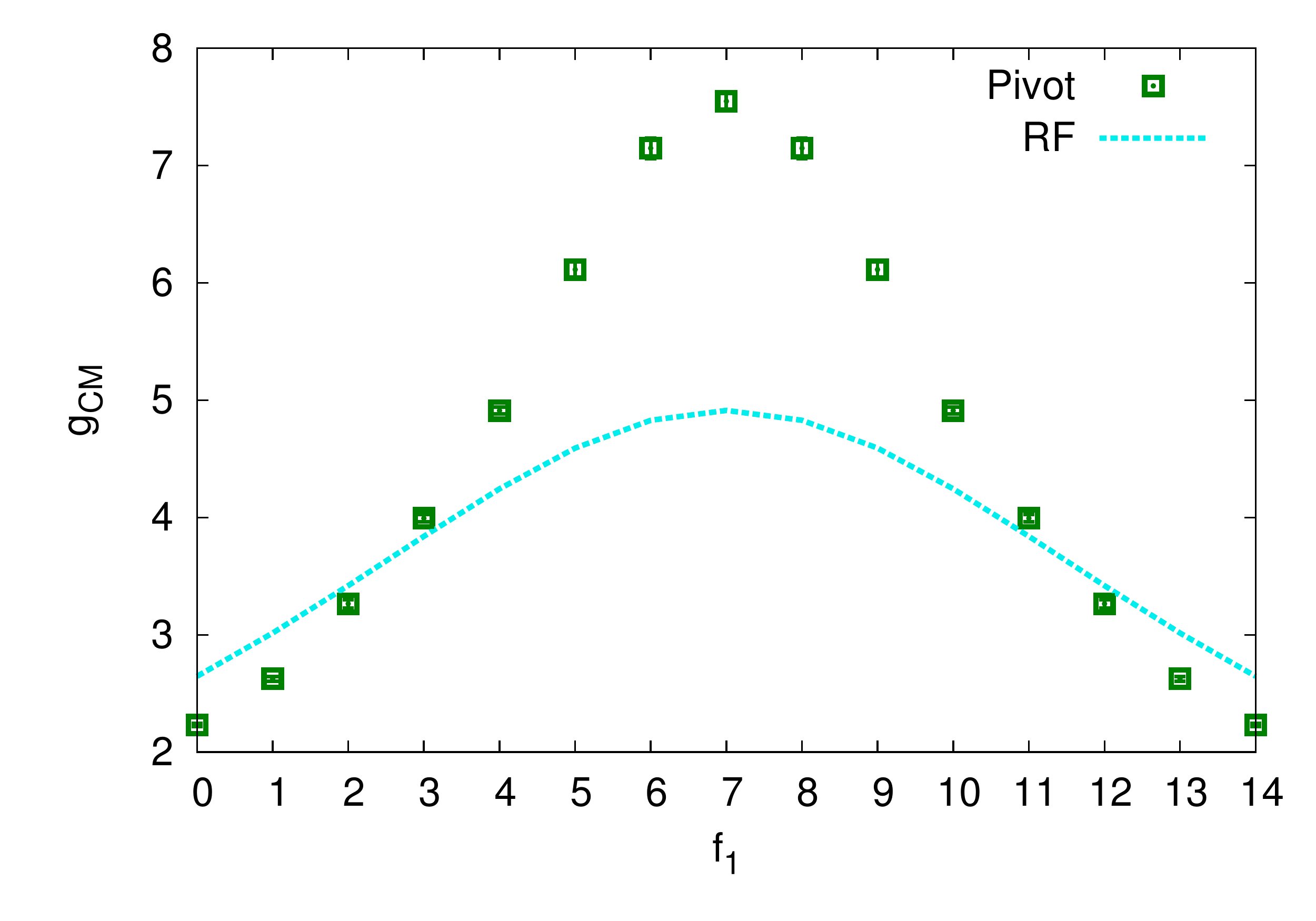}
		\caption{(Colour online) Universal ratio $g_{CM}$ (\ref{CMratio}), that characterizes the average shift of the center of mass of the pom-pom structure from the position of its central monomer, evaluated with respect to such deviation for a linear chain with the equivalent molecular mass. Analytic result in the Gaussian approximation is shown via dashed cyan line, Monte Carlo results using the pivot algorithm --- via green squares.
		\label{CMchainpom}
		}
	\end{center}
\end{figure}

Hydrodynamic properties of the polymer solutions are often studied by replacing each polymer molecule by an effective soft convex body (e.g., the sphere or equivalent ellipsoid), and, consequently, the center-center effective interactions between such bodies are examined~\cite{Bolhuis01,Murat98,Likos98}.
Thus, the center of mass (CM) of each polymer appears in a natural way, as well as the distribution of monomers around it~\cite{Stockmayer53,Krakoviack02}.
For the evaluation of the quantities presented in this subsection, the random-flight model (section~\ref{flight}) approach is the most convenient. The CM of a molecule is defined, therefore, as:
\begin{eqnarray}
&&\vvec{R}_{CM}=\frac{1}{N(f_1+f_2+1)}\left[ \sum_{n=1}^{N/2}\left(\vvec{R}_{{\rm cl}n}+\vvec{R}_{{\rm cr}n}\right)+\sum_{n=N/2}^{N+N/2}\left(\sum_{i=1}^{f_1}\vvec{R}^{i}_{{\rm pl}n}+\sum_{i=1}^{f_2}\vvec{R}^{i}_{{\rm pr}n} \right)\right ].
\end{eqnarray} 
Expressing the position vectors  as sums of bond vectors [equations~(\ref{centralR})--(\ref{pomR2})], we have
\begin{align}
\vvec{R}_{CM}&=\frac{1}{N(f_1+f_2+1)}\left[ \sum_{n=1}^{N/2}\left (  \sum_{k=1}^n \boldsymbol{\sigma}_{{\rm cl}k}+ \sum_{k=1}^n \boldsymbol{\sigma}_{{\rm  cr}_k}\right)  \right. +\sum_{n=N/2}^{N+N/2} \sum_{i=1}^{f_1}   \left(  \sum_{k=1}^{N/2} \boldsymbol{\sigma}_{{\rm  cl}k}+ \sum_{k=1}^n \boldsymbol{\sigma}^{i}_{{\rm  pl}k}\right)\nonumber\\
&+\left. \sum_{n=N/2}^{N+N/2}    \sum_{i=1}^{f_2}   \left( \sum_{k=1}^{N/2} \boldsymbol{\sigma}_{{\rm cr}k}+ \sum_{k=1}^n \boldsymbol{\sigma}^{i}_{{\rm pr}k}\right)\right]=\frac{1}{N(f_1+f_2+1)}\left\{ \sum_{n=1}^{N/2}\left(  \boldsymbol{\sigma}_{{\rm cl}n}+ \boldsymbol{\sigma}_{{\rm  cr}n}\right) \left(\frac{N}{2}-n\right)  \right. \nonumber\\
&+ \sum_{i=1}^{f_1} \left[  N \sum_{n=1}^{N/2} \boldsymbol{\sigma}_{{\rm cl}n}   +    \sum_{n=1}^N \boldsymbol{\sigma}^{i}_{{\rm  pl}n}(N-n)\right]+\left. \sum_{i=1}^{f_2} \left[  N \sum_{n=1}^{N/2} \boldsymbol{\sigma}_{{\rm cr}n}   +    \sum_{n=1}^N \boldsymbol{\sigma}^{i}_{{\rm  pr}n}(N-n)\right] \right\}.
\end{align} 
Taking into account relations (\ref{sigma1}) and (\ref{sigma2}),
we obtain an expression for the mean-squared location of the center of mass:
\begin{eqnarray}
 \langle \vvec{R}_{CM}^2 \rangle_{\rm pom-pom} =\frac{N\left[ 1+7(f_1+f_2)+6(f_1^2+f_2^2)\right]}{12(f_1+f_2+1)^2}.
\end{eqnarray}
For the center of mass of a single chain of the same total length with a reference point in the middle of a chain, we have:
\begin{eqnarray}
 \langle \vvec{R}_{CM}^2 \rangle_{\rm chain} =\frac{N(1+f_1+f_2)}{12 }.
 \label{CMchain}
\end{eqnarray}
Thus, we can introduce the ratio:
\begin{equation}
g_{CM}=\frac{\langle \vvec{R}_{CM}^2 \rangle_{\rm chain} }{\langle \vvec{R}_{CM}^2\rangle_{\rm pom-pom} },
\label{CMratio}
\end{equation} 
which describes the impact of the presence of  side stars (see figure \ref{CMchainpom}) on the shift of the CM position of complex polymer structure as compared with that of linear chain as given by equation~(\ref{CMchain}). This quantity may serve as yet another characteristic of the asymmetry of the pom-pom structure. In the Gaussian approximation, we have: 
\begin{equation}
g_{CM}=\frac{(f_1+f_2+1)^3}{1+7(f_1+f_2)+6(f_1^2+f_2^2)}. \label{CMratio1}
\end{equation} 
\begin{figure}[t]
	\begin{center}
		\includegraphics[width=60mm,angle=270]{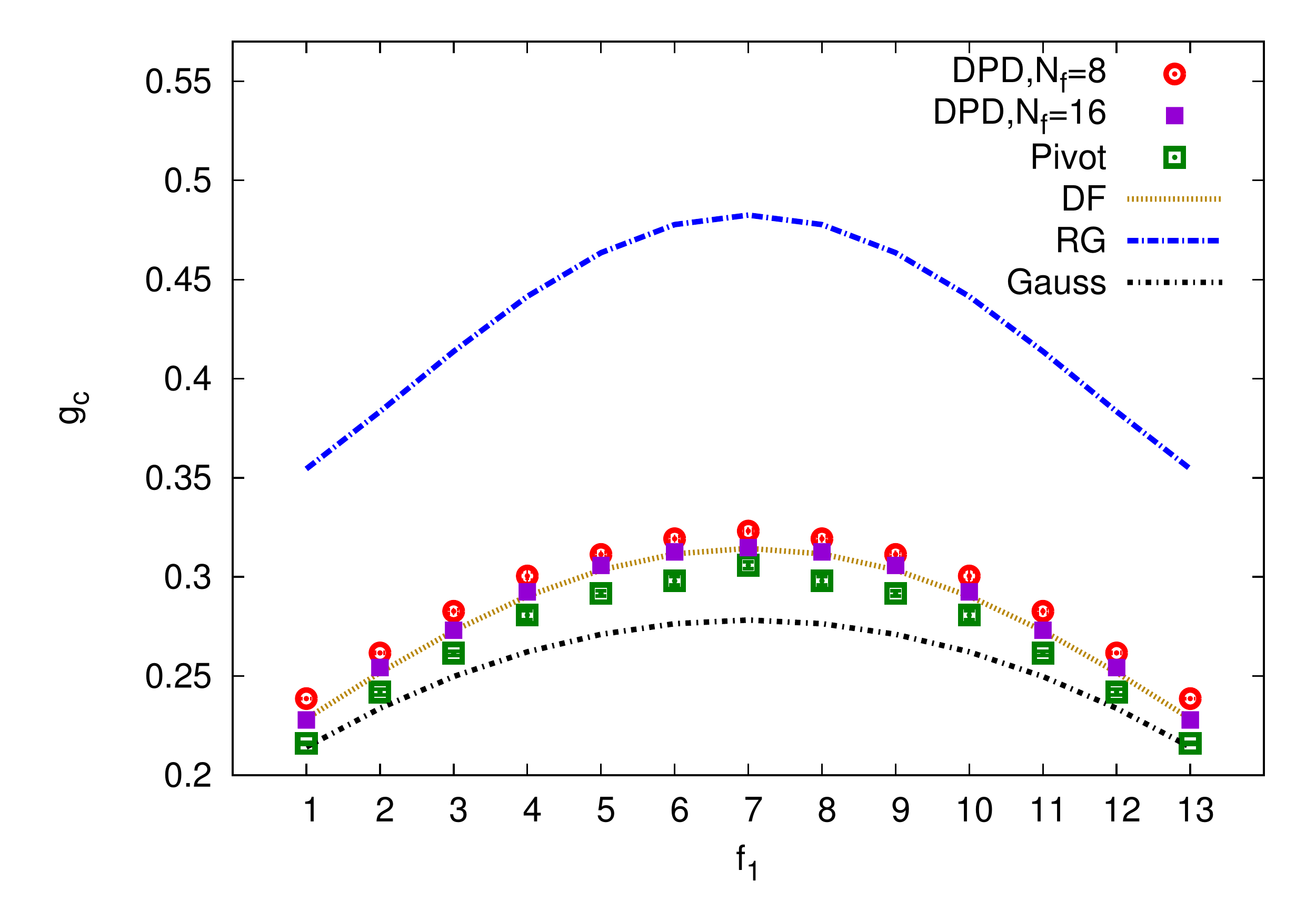}
		\includegraphics[width=60mm,angle=270]{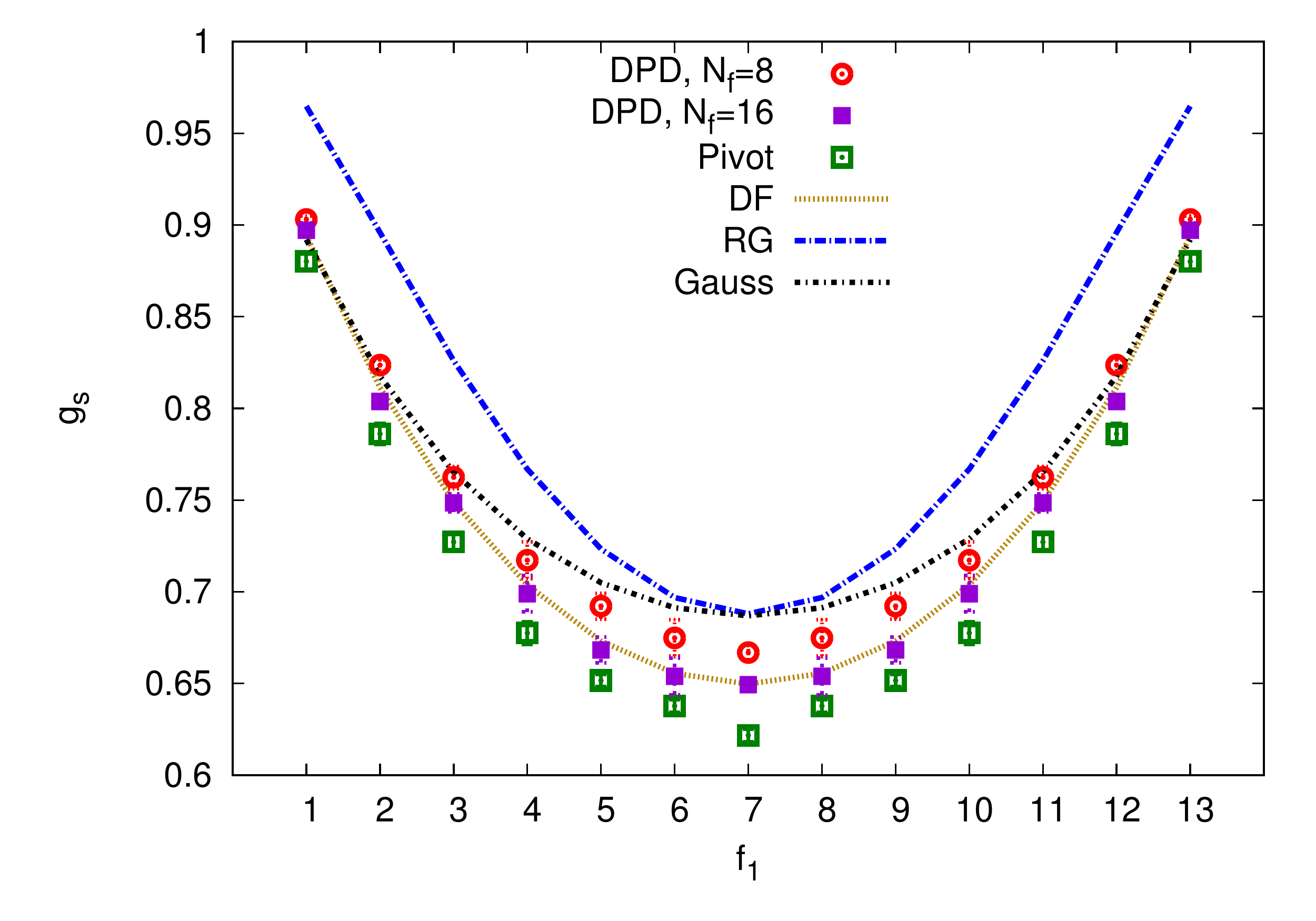}
		\caption{ \label{fig:6} (Colour online)  Size ratios $g_c$ (\ref{gc}) (top) and $g_s$ (\ref{gs}) (bottom) vs branching parameter $f_1$. The total number of side chains is fixed at $F=f_1+f_2+1=15$. Red open disks and magenta squares are the results of the DPD simulations at $N_f=8$ and $N_f=16$, respectively, green squares --- the results of the MC simulations using the pivot algorithm. Blue curves marked ``RG'' represent expressions (\ref{gc1}) and (\ref{gs1}) evaluated at fixed point (\ref{FPP}) for $\epsilon=1(d=3)$. Yellow dash line shows the results of the Douglas--Freed approximation and a black dot-dash line represents the Gaussian model. }
	\end{center}
\end{figure}
\begin{figure}[t]
	\begin{center}
		\includegraphics[width=60mm,angle=270]{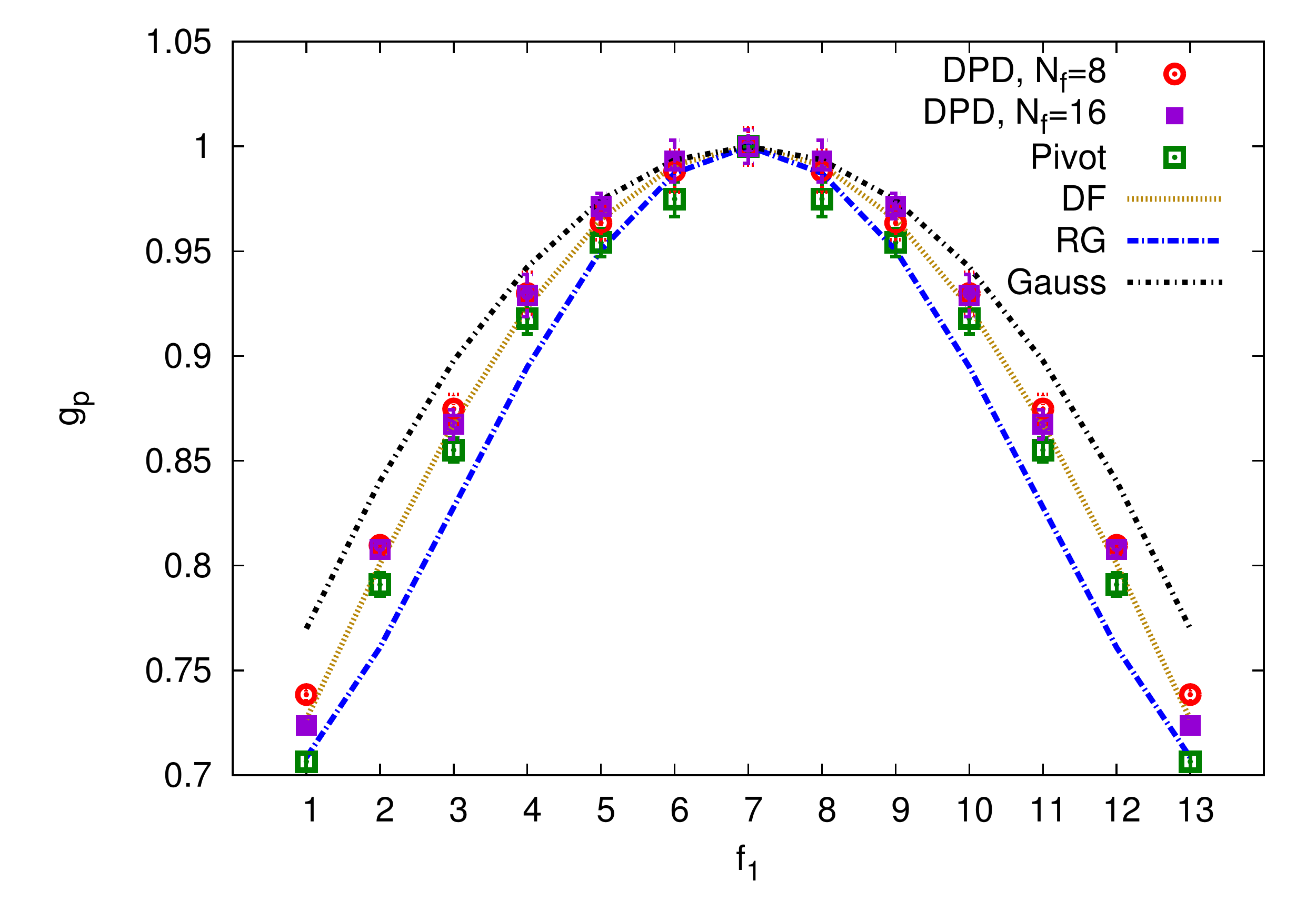}
		\includegraphics[width=60mm,angle=270]{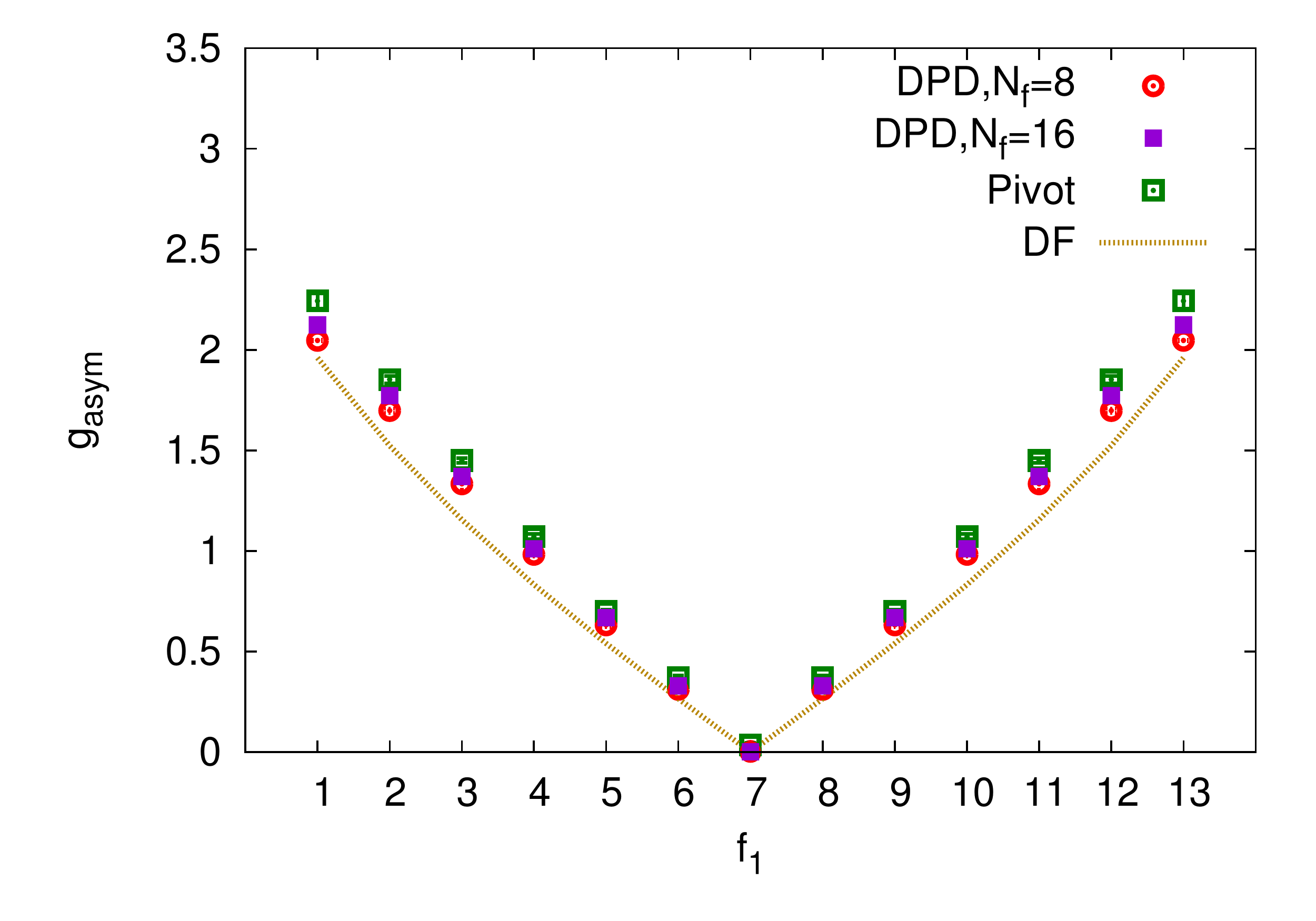}
		\caption{ \label{fig:7} (Colour online) The same as in figure~\ref{fig:6} but for size ratio $g_p$ (\ref{gp}) (top) and asymmetry factor~$g_{\rm{asym}}$~(\ref{gasym})~(bottom).}
	\end{center}
\end{figure}
Let us introduce the distance of each $n$-th monomer from the CM
\begin{equation}
 \vvec{R}_{nCM}=\vvec{R}_n -\vvec{R}_{CM}.      
\end{equation}
Applying the scheme introduced above, we obtain the analytic expressions for these distances for the monomers belonging to a central chain and these from the left and right branches, respectively:
\begin{eqnarray}
&&\langle \vvec{R}^2_{n{\rm cl}CM}  \rangle=n\left[ 1- \frac{(2f_1+1)}{f_1+f_2+1}+\frac{n}{(f_1+f_2+1)N}\right]+ \langle \vvec{R}_{CM}^2\rangle,\label{rc1} \\
&&\langle \vvec{R}^2_{n{\rm cr}CM}\rangle=n\left[ 1- \frac{(2f_2+1)}{f_1+f_2+1}+\frac{n}{(f_1+f_2+1)N}\right]+ \langle \vvec{R}_{CM}^2\rangle,\label{rc2}\\
&&\langle \vvec{R}^2_{n{\rm pl}CM}\rangle=n\left[ 1-  \frac{3}{(f_1+f_2+1)}+\frac{n}{(f_1+f_2+1)N}  \right]+\frac{N(1-f_1)}{(f_1+f_2+1)}+ \langle \vvec{R}_{CM}^2\rangle,\label{rcrightpom} \\
&&\langle \vvec{R}^2_{n{\rm pr}CM}\rangle=n\left[ 1-  \frac{3}{(f_1+f_2+1)}+\frac{n}{(f_1+f_2+1)N}  \right] +\frac{N(1-f_2)}{(f_1+f_2+1)}+ \langle \vvec{R}_{CM}^2\rangle.\label{rcleftpom}
 \end{eqnarray}


\subsection{Numeric results and discussion}
The main focus of this study are the theoretic results for the shape characteristics of asymmetric pom-pom polymers, whereas the numeric studies, performed by means of the DPD and MC simulations, were conducted to support these theoretical findings. To this end, a set of size ratios, defined above, were obtained by means of the above mentioned numeric approaches. This set includes: the ratio between the gyration radius of asymmetrical pom-pom to that of a linear chain $g_c$ (\ref{gc}), of a star-like polymer (\ref{gs}) and of a symmetrical pom-pom polymer (\ref{gp}). These results, alongside with the theoretical findings, are shown in figures~\ref{fig:6} and \ref{fig:7}. Red open discs and magenta squares represent the DPD simulations with $N_f=8$ and $N_f=16$, respectively, green squares represent the MC study using a pivot algorithm. Yellow dash line displays the Douglas--Freed approximation, blue dash line: the first order (one loop) renormalization group results, and a black dot-dash line: these obtained within the Gaussian model. 

The results for the $g_c$ ratio (\ref{gc}) are shown in the top frame of figure~\ref{fig:6}. \emph{First observation} is that $g_c$ is always below $1$. This is a natural consequence of the pom-pom connectivity, which organizes monomers into two star-like bunches, which cannot stretch as much as a single linear chain of the same molecular weight. Hence, the gyration radius of a pom-pom  is always smaller than that of a linear chain. \emph{Second observation} is that the $g_c$ curve has a convex shape and $g_c$ is minimal for the highly asymmetric pom-pom structures, $f_1=1$ and $f_1=13$, since in this case one of the branching points holds a maximum possible number of arms and thus maximizes the spatial confinement effect. \emph{Third observation} is that there is a reasonable agreement between the results of the DPD and MC simulations, despite great difference between soft potential and off-lattice nature of the former and hard potentials and lattice arrangement of monomers in the latter. The Douglas--Freed approximation is found to be in a good agreement with the simulation data, coinciding perfectly with the DPD results obtained at the arms length of $N_f=16$. This shows that such arms length is long enough to mimic the ``long arms limit'' assumed in the theory, and that $g_c$ clearly demonstrates a strong universality feature. \emph{Fourth observation} is about the Gaussian result for $g_c$, which is found below both simulation results. Switching off the excluded volume effects (the case of the Gaussian model), reduces the denominator of $g_c$: $R^2_{g,{\rm chain}}\sim N^{0.5}$ as compared to its value $R^2_{g,{\rm chain}}\sim N^{0.59}$ for the case with excluded volume effects (as it is in both simulations here). As far as the ratio $g_c$ decreases in the case of the Gaussian model, the nominator $R^2_{g,{\rm pom-pom}}$ in the Gaussian approximation must decrease even more compared to the case of simulations, indicating a stronger effect of ``compactization'' of a pom-pom structure compared to the linear chain when the excluded volume effects are switched off. Finally, the \emph{fifth observation} is that the first order renormalization group result overestimates $g_c$ essentially in the whole interval of $f_1$. This situation is somewhat typical of the renormalization group findings for other properties of interest, and a much better qualitative agreement is typically expected when the second order terms are included. 

Similar analysis can be conducted for other ratios. Namely, for the ratio $g_s$ (\ref{gs}), shown in the bottom frame of figure~\ref{fig:6}, one sees that $g_s<1$, indicating that a star polymer is always more compact than a pom-pom polymer, which is obvious, because the latter contains two starts separated by a backbone and spreads wider in space. In limit cases of $f_1=1$ and $f_1=13$, a pom-pom structure reduces to a single star with one of its arms of a length $2N_f$ and, therefore, $g_s\approx 1$. The difference in the $g_s$ values obtained for these two molecular structures is the most pronounced for a symmetric case, $f_1=f_2=7$, leading to a concave shape for the $g_s$ curve. The third and fourth observations made for $g_c$ (see above) also hold for the case of $g_s$. However, the first order renormalization group results, albeit being closer to the Gaussian results than in the case of $g_c$, are shifted away from the simulation data.

Top frame of figure~\ref{fig:6} shows the $g_p$ ratio, which characterizes the effect of asymmetry on the squared gyration radius of a pom-pom structure. According to the definition (\ref{gp}), $g_p=1$ for the symmetric case of $f_1=f_2=7$. In the case of a maximum asymmetry, $f_1=1$ and $f_1=13$, $g_p$ drops down to about~$0.7$, reflecting reduction of a pom-pom structure to that of a single star, which is a much more compact structure. The third, fourth and fifth observations made for $g_c$ also hold here. 

As a direct convenient measure of the asymmetry of a pom-pom structure, we introduced the asymmetry factor $g_{\rm{asym}}$ (\ref{gasym}). This is shown in the bottom frame of figure~\ref{fig:6}. One can see that it grows almost linearly when $f_1$ moves from the $f_1=7$ value towards both sides of the plot, and the results of the Douglas--Freed approximation agree reasonably well with all the simulation data, supporting the same trend that was observed for $g_c$, $g_s$ and $g_p$ ratios. 

\section{Conclusions}\label{conclusions}

In the present paper, we combined analytical studies and two types of computer simulations to evaluate the set of universal characteristics of the asymmetric pom-pom polymers. Such polymers are characterized by the functionality of their branching points, $f_1$ and $f_2=F-1-f_1$, where $F=f_1+1+f_2=15$ is the total number of arms, each of the length $N_f$. Theoretical approach is based on direct polymer renormalization of the Edwards continuous chain model and a random-flight model. Computer simulations employ both the off-lattice dissipative particle dynamics approach with explicit good solvent, and the Monte Carlo approach using a pivot algorithm on a lattice model of polymer assuming an implicit solvent.

As expected from its molecular architecture, characteristic size of a pom-pom structure generally fits in between that for a linear chain and of a star. To characterize these findings quantitatively at various levels of the pom-pom asymmetry, we introduced  a set of universal dimensionless ratios. These are: $g_c$ (size ratio of a pom-pom relative to that of a linear chain), $g_s$ (the same relative to a star polymer) and~$g_p$ (the same relative to a symmetric pom-pom). The molecular mass of all polymer molecules is set the same. The asymmetry of a pom-pom polymer is proposed to be characterised also by the factor $g_{\rm{asym}}$, which is a combination of the average squared distances $\langle r^2_1 \rangle$ and $\langle r^2_2 \rangle$ of the monomers from the first and second branching points, respectively. It is defined in such a way that it is equal to $0$ for a symmetric case ($f_1=f_2$) and reaches its maximum value for the most asymmetric cases, e.g., when $f_1=1$ and $f_2=F-2$. We also introduced the ratio $g_{CM}$, which provides a quantitative estimate of the shift of the center of mass of the central backbone caused by the presence of side stars, and serves as an additional characteristic of the asymmetry of a pom-pom structure.

Various simulation approaches applied by us, despite their differences in a description of both a polymer and a solvent, provide very close results for all size ratios as functions of $f_1$ and $f_2$. This is indicative of universality features for all ratios in a broad sense, meaning that they depend solely on the details of molecular architecture rather than on the chemical particulars. We also found a clear deviation of the numeric results from those obtained within the Gaussian approximation, showing the nontrivial role of excluded volume effects. The first order (one loop approximation) renormalization group results provide a nice correction to the Gaussian case, but in most cases it would be worthwhile to proceed to higher orders of perturbation theory terms  as well. The latter is, however, quite complicated and cumbersome and is extremely time consuming. Instead, a good practical solution is to use the so-called Douglas--Freed approximation~\cite{Douglas84} which accounts for a relevant subset of the second order contributions. The effectiveness of this approximation is once again shown in this work, since both simulation results agree well with it, especially the DPD simulations with longer arms, $N_f=16$, indicating that the model reaches a ``long chain limit'' in this case.

It is important to point out that some
marked differences between the simulations and theory are observed within this study. Those differences may be related to limitations of both analytical and numerical methods. For example, lower values for some ratios obtained by means of the MC simulation as compared with analytical ones may be caused by more limited conformational space of a lattice model, whereas higher values in the case of the DPD simulations at $N_f=8$ are side-products of considering chains, which are too short. 

Results provided in this study may be of use for further research of the pecularities of complex polymers in solution. Numerical studies may be extended to deal with semi-dilute or dense solutions, micelles formation, etc. 

\section*{ Acknowledgements}

The authors would like to acknowledge support by the grant programs 6541230  (fundamental research) of the National Academy of Sciences of Ukraine ``Self-assembly and formation of structures in Soft Matter: analytical approaches and computer simulations'' (reg. no. 0121U110784) and ``Emergence in Soft Matter: models and applications'' (reg. no. 0122U002354). 

K.~H. would like to acknowledge the support from the National
Science Center, Poland (Grant No.~2018/30/E/ST3/00428).

V.~B. is grateful for support from the U.S. National Academy of Sciences (NAS) and the Polish Academy of Sciences (PAS) to scientists from Ukraine.

We thank the Ukrainian Army for the possibility to perform this research work.
\newpage
\section*{Appendix}

\renewcommand{\theequation}{A.\arabic{equation}}

\subsection*{A list of expressions}

An expression for $C_g(f_1,f_2,d)$ from the expression (\ref{RG}):
\begin{align}
-C_g(f_1,f_2,d)&=\frac{2}{3f_1^2+12f_1f_2+3f_2^2+4f_1+4f_2+1}\left\{\frac{12f_1^2f_2^2(2^{2-{d}/{2}}-3^{1-{d}/{2}}-1)}{(d-2)d}\right.\nonumber\\
&+\frac{2^{3-{d}/{2}}f_1f_2[4d(14-d)(d^2-14d+184)-38400]}{(d-10)(d-6)d(d-8)(d-2)(d-4)}\nonumber\\
&+\frac{24(f_1+f_2)(d^3-18d^2+40d-448)f_2f_12^{-{d}/{2}}}{(d-2)d(d-6)(d-4)(d-8)} \nonumber\\
&-\frac{72(f_1+f_2)(d^3-18d^2-4d-192)f_2f_13^{-{d}/{2}}}{(d-2)d(d-6)(d-4)(d-8)}\nonumber\\
&-\frac{9f_1f_23^{-{d}/{2}}(7d^4-196d^3+1124d^2+16d-38400)}{(d-10)(d-6)d(d-8)(d-2)(d-4)}\nonumber\\
&+\frac{4(d-12)(d^2-6d+32)(f_1+f_2)(f_1^2-f_1f_2+f_2^2-1)}{(d-2)d(d-6)(d-4)(d-8)}\nonumber\\
&+\frac{(d^2-26d+136)(f_1^2+f_2^2)}{(d-4)(d-10)(d-6)(d-8)}+\frac{[12(f_1+f_2)](d^2-10d+32)f_1f_2}{d(d-2)(d-4)(d-6)}\nonumber\\
&-\frac{f_1f_2(13d^4-364d^3+3788d^2-18512d+38400)}{(d-10)(d-6)d(d-8)(d-2)(d-4)}\nonumber\\
&+\frac{-d^5+32d^4-300d^3+1024d^2-1088d}{(d-10)(d-4)^2(d-2)d(d-6)(d-8)} \nonumber\\
&-\frac{2^{9-{d}/{2}}f_1(f_1+1)(47df_1-13d-120f_1+120)}{(d-10)(d-4)^2(d-2)d(d-6)(d-8)}\nonumber\\
&+\frac{2^{9-{d}/{2}}f_2(f_2+1)(47df_2-13d-120f_2+120)}{(d-10)(d-4)^2(d-2)d(d-6)(d-8)}\nonumber\\
&\left.+
\frac{[2^{3-{d}/{2}}d^2f_2(f_2+1)^2+f_1(f_1+1)^2]}{(d-10)(d-4)^2(d-2)d(d-6)(d-8)}(d^3-32d^2+300d-1024)\right\}.\label{A_rg}
\end{align}
An expression for $A_g(f_1,f_2)$ from the expression (\ref{chain}--\ref{ppa}):
\begin{eqnarray}
A_g(f_1,f_2)&&=-\frac{13}{12}-\left(3f_1^2+12f_1f_2+3f_2^2+4f_1+4f_2+1\right)^{-1}\nonumber\\
&&\times\biggl\{\frac{f_1f_2}{2}(2f_1f_2-41f_1-41f_2+28)+\frac{13}{6}\biggr.-\frac{13(f_2+1)}{6}(3f_2^2+1)\nonumber\\
&&-\frac{13(f_1+1)}{6}(3f_1^2+1)
+4\ln2\left[f_2(f_2+1)(3f_2-2)+f_1(f_1+1)(3f_1-2)\right]\nonumber\\
&&\biggl.-16\ln2f_1f_2(3f_2+3f_1-5)+
27\ln3f_1f_2(2f_2+2f_1-3)\biggr\}.\label{A_A_g}
\end{eqnarray}

An $\epsilon$-expressions for $\langle {r_1^2}\rangle$ and $\langle {r_2^2}\rangle$ introduced in (\ref{r_1}) and (\ref{r_2}):

\begin{align}
{\langle {r_1^2}\rangle} &=\frac{dL(f_1+1+3f_2)}{2 F} \biggl\{1+u_0\biggl[\frac{2}{\epsilon}+\frac{1}{12(f_1+1+3f_2)}\biggr.\biggr.(12\ln2(2f_1^2-20f_1f_2+2f_1+7f_2)\nonumber\\ \label{A_C1}
&\biggl.\biggl.+216\ln3f_1f_2-13f_1^2-40f_1f_2-27f_1-52f_2-14-12f_1f_2^2)\biggr]\biggr\},
\end{align}

\begin{align}
{\langle {r^2_2}\rangle} &=\frac{dL(f_2+1+3f_1)}{2 F}\biggl\{1+u_0\biggl[\frac{2}{\epsilon}+\frac{1}{12(f_2+1+3f_1)}\biggr.\biggr.(12\ln2(2f_2^2-20f_1f_2+2f_2+7f_1)\nonumber\\ \label{A_C2}
&\biggl.\biggl.+216\ln3f_1f_2-13f_2^2-40f_1f_2-27f_2-52f_1-14-12f_2f_1^2)\biggr]\biggr\}.
\end{align}


\newpage
\ukrainianpart

\title{Властивості форми асиметричних пом-пом полімерів у хорошому розчиннику}

\author[Х. Гайдуківська, O. Калюжний, В. Блавацька, Я. Ільницький]
{Х. Гайдуківська\refaddr{label1,label2},
 O. Калюжний\refaddr{label1},
 В. Блавацька\refaddr{label1}, 
 Я. Ільницький\refaddr{label1,label3}}

\addresses{
\addr{label1} Інститут фізики конденсованих систем Національної академії наук України, вул. Свєнціцького, 1, 79011, Львів, Україна
\addr{label2} Інститут Фізики, Сілезький університет, вул. Першого полку піхоти, 75, 41-500 Хоржув, Польща
\addr{label3} Діоскурі центр фізики і хімії бактерій,
Інститут фізичної хімії, Польська академія наук, 01-224 Варшава, Польща
\addr{label3} Національний університет ``Львівська політехніка'', вул. Бандери, 12, 79000, Львів, Україна
}

\makeukrtitle

\begin{abstract}
\tolerance=3000%
Ця стаття є продовженням нашого попереднього дослідження [K. Haydukivska et al., J. Mol. Liq., 2021, \textbf{328}, 115456] складних молекул з двома центрами галуження на кінцях ланцюжка основи з $f_1$ та $f_2$ боковими ланцюжками прикріпленими до них, які відомі як пом-пом полімери. Тут ми аналізуємо асиметричний випадок, $f_1 \neq f_2$, використовуючи як аналітичний підхід на основі прямого полімерного перенормування, так і числьне моделювання з використанням методів дисипативної динаміки та Монте-Карло. Ми досліджуємо вплив молекулярної асиметрії на усереднену конформацію, розглядаючи випадок безмежного розведення в хорошому розчиннику. Були отримані кількісні оцінки для низки універсальних характеристик розміру та форми для досліжуваних молекул та їх складових гілок, як функції $f_1$ та $f_2$. До прикладу, ми оцінюємо розмірне співвідношення для радіусів гірації симетричної та асиметричої топологій пом-пом однакової молекулярної маси, та кількісно показуємо ефект зросту ефективного розміру молекули, спричиненого ассиметрією. Ми також вводимо і аналізуємо фактор асиметрії та оцінюємо зміщення цент\-ру мас через  присутність бокових зірок, що може слугувати ще однією характеристикою асиметричності полімерної структури.
\end{abstract}

\end{document}